\documentclass{elsart}
\usepackage{color,setspace,times}
\usepackage{amssymb,amsmath,graphicx,color,rotating,subfigure,url}
\usepackage{lineno}

\journal{Physica A} %% change this journal name and put the correct one

\begin{document}
\begin{frontmatter}

\title{Intraday pattern in bid-ask spreads and its power-law relaxation for
Chinese A-share stocks}
\author[BS,SS]{Xiao-Hui Ni},
\author[BS,SS,CES,RCSE]{Wei-Xing Zhou \corauthref{cor}}
\corauth[cor]{Corresponding author. Address: 130 Meilong Road, P.O.
Box 114, School of Business, East China University of Science and
Technology, Shanghai 200237, China, Phone: +86 21 64253634, Fax: +86
21 64253152.}
\ead{wxzhou@ecust.edu.cn} %

\address[BS]{School of Business, East China University of Science and Technology, Shanghai 200237, China}
\address[SS]{School of Science, East China University of Science and Technology, Shanghai 200237, China}
\address[CES]{Center for Econophysics Studies, East China University of Science and Technology, Shanghai 200237, China}
\address[RCSE]{Research Center of Systems Engineering, East China University of Science and Technology, Shanghai 200237, China}

\begin{abstract}
We use high-frequency data of 1364 Chinese A-share stocks traded on
the Shanghai Stock Exchange and Shenzhen Stock Exchange to
investigate the intraday patterns in the bid-ask spreads. The daily
periodicity in the spread time series is confirmed by Lomb analysis
and the intraday bid-ask spreads are found to exhibit $L$-shaped
pattern with idiosyncratic fine structure. The intraday spread of
individual stocks relaxes as a power law within the first hour of
the continuous double auction from 9:30AM to 10:30AM with exponents
$\beta_{\rm{SHSE}}=0.20\pm0.067$ for the Shanghai market and
$\beta_{\rm{SZSE}}=0.19\pm0.069$ for the Shenzhen market. The
power-law relaxation exponent $\beta$ of individual stocks is
roughly normally distributed. There is evidence showing that the
accumulation of information widening the spread is an endogenous
process.
\end{abstract}


\end{frontmatter}

\section{Introduction}

In most modern financial markets, the intraday pattern exists
extensively in many financial variables
\cite{Wood-McInish-Ord-1985-JF,Harris-1986-JFE,Admati-Pfleiderer-1988-RFS},
including the bid-ask spread \cite{McInish-Wood-1992-JF}. The
periodic pattern has significance impact on the detection and
understanding of long memory in time series
\cite{Hu-Ivanov-Chen-Carpena-Stanley-2001-PRE}. To the best of our
knowledge, detailed investigation of intraday patterns in the
bid-ask spreads of Chinese stocks is rare
\cite{Gu-Chen-Zhou-2007-EPJB}.

Empirical studies have examined the intraday pattern in bid-ask
spreads for various types of stock markets. It has been documented
that the intraday width of bid-ask spreads for New York Stock
Exchange (NYSE) stocks follows a U-shaped pattern, where the spreads
are the widest immediately after the opening of the market and right
before the closure
\cite{Lee-Mucklow-Ready-1991-RFS,Brock-Kleidon-1992-JEDC,McInish-Wood-1992-JF}.
Some researchers studied two samples of National Association of
Securities Dealers Automated Quotation (NASDAQ) stocks and related
the results to the institutional features of the dealer market
\cite{Chan-Christie-Schultz-1995-JB}. They found that, in contrast
to the U-shaped pattern for NYSE stocks, the bid-ask spreads for
NASDAQ securities are relatively stable throughout the day but
narrow significantly during the final hour of trading. These results
are similar to the declining intraday spreads for firms on the
London Stock Exchange during mandatory trading hours
\cite{Werner-Kleidon-1996-RFS}.

Despite of the extreme importance of the bid-ask spreads in the
study of market microstructure, the intraday pattern in spreads of
Chinese stocks has not attracted much attention. The Chinese stock
market operates as an order-driven system, which again account for
only a small part in literature of microstructure theory. The
Chinese stocks are traded on the Shanghai Stock Exchange (SHSE) and
the Shenzhen Stock Exchange (SZSE). The market opens at 9:15AM with
call auctions till 9:25AM and then enters a five-minute cooling
period. The stocks are traded based on continuous double auction
from 9:30AM to 11:30AM and from 13:00PM to 15:00PM. The trading
pauses from 11:30AM to 13:00PM. The intraday spreads of Chinese
stocks thus exhibit different fine structures.

In this work, we investigate the intraday pattern in bid-ask spreads
of 1364 Chinese A-share stocks\footnote{A shares are common stocks
issued by mainland Chinese companies, subscribed and traded in
Chinese RMB, listed in mainland Chinese stock exchanges, bought and
sold by Chinese nationals. A-share market was launched in 1990.}.
The data sets are described in Sec.~\ref{s1:dataset}. We confirm in
Sec.~\ref{s2:Lomb} that the spread time series has a daily
periodicity and show the intraday patterns in bid-ask spreads in
Sec.~\ref{s2:BAS:intraday}. The wide bid-ask spread after the
cooling period implies the cumulation of personal interpretation of
public and private information received before continuous auction.
In the framework of complex systems theory, the relaxation dynamics
of spread contain crucial information about the cumulation process
during the cooling period. We thus investigate in detail the
relaxation behavior of the spreads when entering the continuous
double auction in Sec.~\ref{s1:PLs}. Section \ref{s1:conclusion}
summarizes and provides a brief discussion.

\section{Data sets}
\label{s1:dataset}

\subsection{Description and preprocessing of the raw data}
\label{s2:data_introduction}

We utilize a nice high-frequency database containing 829 stocks
traded on the SHSE and 535 stocks on the SZSE in the Chinese A-share
market covering the whole year of 2005. The data were recorded based
on the market quotes disposed to all traders in every six to eight
seconds, which are different from the ultra-high-frequency data
reconstructed from the limit-order book. Each datum is time stamped
to the nearest second at which one transaction occurs. Therefore,
for each stock $i$, the raw data of interest contain two time
series, $a_i(t)$ for the best ask prices and $b_i(t)$ for the best
bid (or offer) prices, where $t$ is the unevenly spaced calendar
time. There are totally 73,187,642 entries for the SHSE stocks and
48,697,722 for the SZSE stocks.

We adopt the following criteria to avoid possible recording errors.
First, the bid price should not be larger than the ask price and the
data at moment $t$ where $a_i(t)\leqslant b_i(t)$ are removed.
Second, all the data in a given trading day are discarded if the
trading frequency is anomalously low. Third, the data stamped with
times outside the continuous double auction (9:30AM to 11:30AM and
13:00PM to 15:00PM) are eliminated. Fourth, we segment each trading
day into eight 30-minute intervals. The whole data of a trading day
are excluded in case any one of the eight intervals is empty such
that the possible intraday pattern in the spread for the stock is
reserved.

\subsection{Constructing 30-second bid-ask spread}
\label{s2:market_time_series_create}

There are different definitions for the bid-ask spread in literature
\cite{Roll-1984-JF,Stoll-1989-JF,Huang-Stoll-1997-RFS,Daniels-Farmer-Gillemot-Iori-Smith-2003-PRL,Farmer-Gillemot-Lillo-Mike-Sen-2004-QF,Plerou-Gopikrishnan-Stanley-2005-PRE,Farmer-Patelli-Zovko-2005-PNAS,Wyart-Bouchaud-Kockelkoren-Potters-Vettorazzo-2006,Cajueiro-Tabak-2007-PA,Mike-Farmer-2007-JEDC},
which nevertheless do not ruin the possible intraday pattern in
spread. The ``point'' definition of spread for stock $i$ at time $t$
is the difference between best bid and ask prices:
\begin{equation}\label{Eq:s:t}
    s_i(t)=a_i(t)-b_i(t)~.
\end{equation}
We stress that $t$ is unevenly spaced calendar time. Each trading
day is evenly divided into 480 time intervals with a length of 30
seconds. The 30-second spread is then defined as the average of the
available spreads within each interval
\begin{equation}\label{Eq:s:1m}
    S_i(t')=\langle{s_i(t)}\rangle, ~~ t\in {\mathbf{T}}(t')
\end{equation}
where ${\mathbf{T}}(t')$ is the $t'$-th 30-second time interval. For
simplicity, $S_i(t')$ is set to be zero if there is no recorded data
in ${\mathbf{T}}(t')$ such that each trading day corresponds to 480
consecutive data points in the time series $S_i(t')$.

\begin{figure}[htb]
\centering
\includegraphics[width=6.5cm]{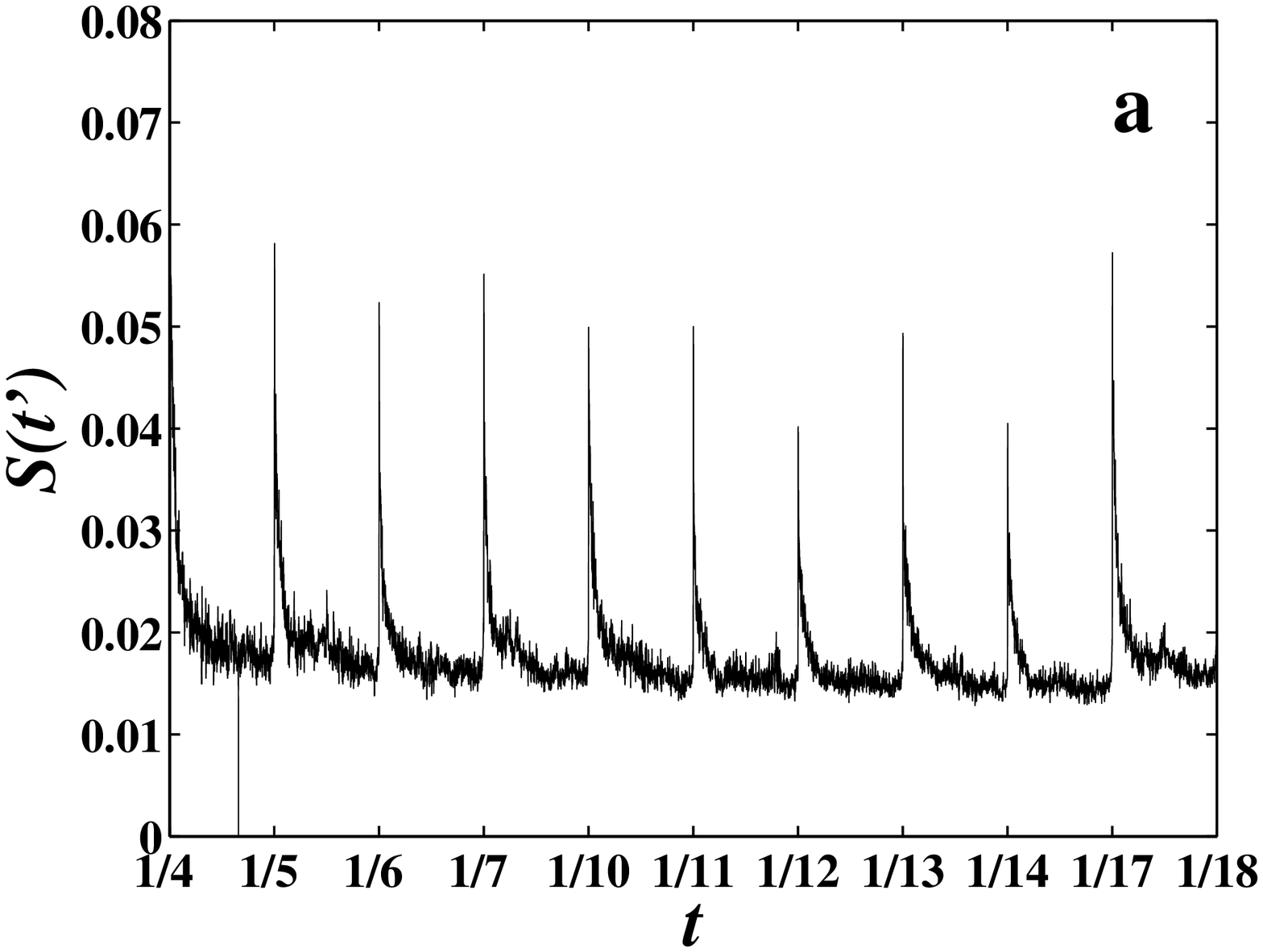}
\includegraphics[width=6.5cm]{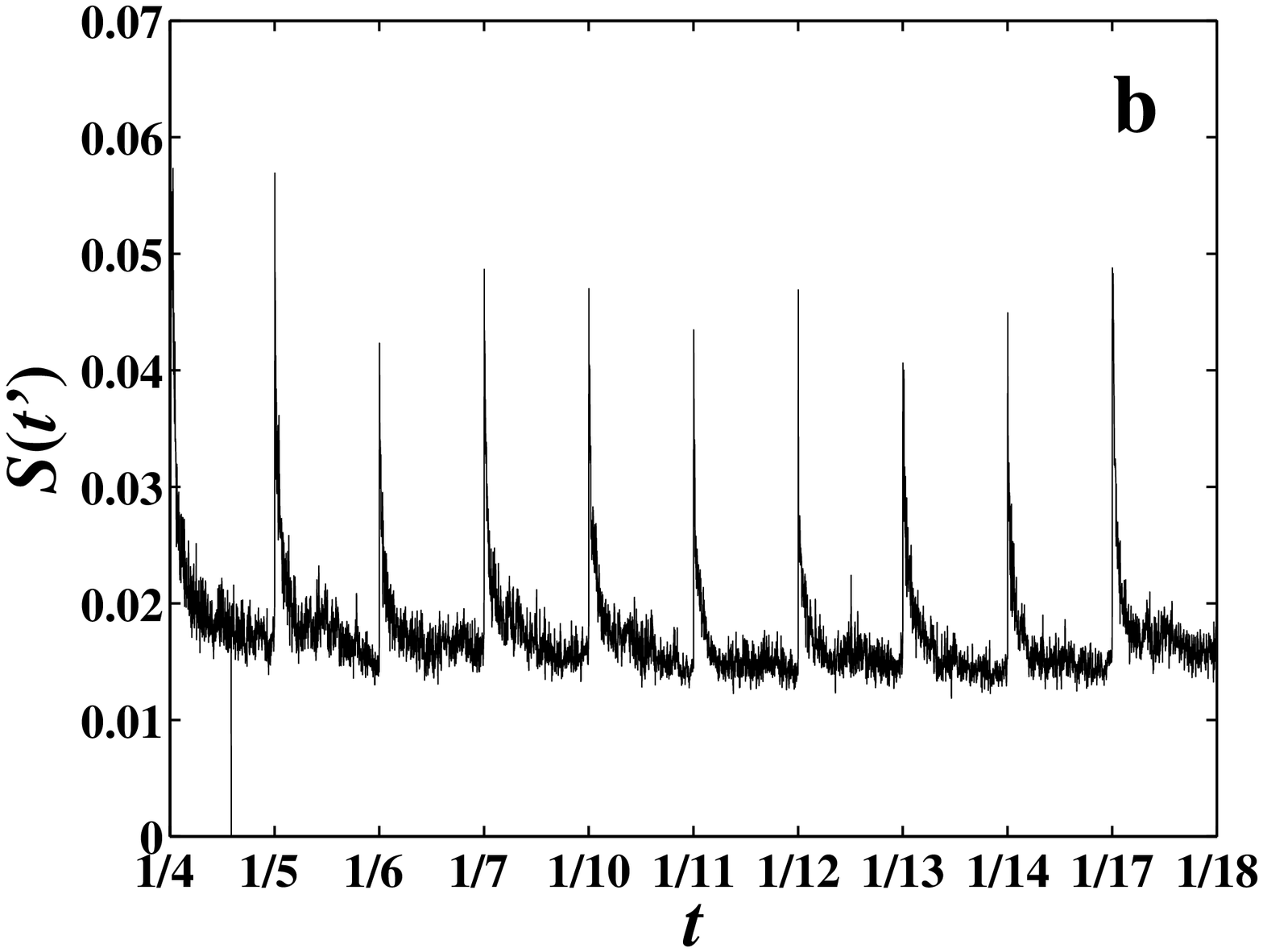}
\caption{\label{Fig:S} Parts of the market-averaged time series of
the 30-second bid-ask spreads: (a) SHSE, (b) SZSE. Intraday pattern
is clearly visible.}
\end{figure}

The market-averaged bid-ask spread is
\begin{equation}\label{Eq:PDF:spreadseries}
    S(t')=\frac{1}{N(t')}\sum_{i\in {\mathbf{\Phi}}(t')}{S_i(t')}~,
\end{equation}
where the set ${\mathbf{\Phi}}(t')=\{j: S_j(t')\neq0\}$ and $N(t')$
is the size of ${\mathbf{\Phi}}(t')$. Figure \ref{Fig:S} shows two
market-averaged time series of the 30-second bid-ask spreads for all
A-shares on the SHSE and SZSE, respectively. We observe that the
spread exhibits daily periodicity and declines during the continuous
double auction.

\section{Characterizing intraday pattern in spread}
\label{s1:IntradayPattern}

\subsection{The daily periodicity}
\label{s2:Lomb}

As a first step, we attempt to confirm daily periodicity in spread
time series observed in Fig.~\ref{Fig:S}. The normalized Lomb power
is used for this purpose
\cite{Lomb-1976-ApSS,Scargle-1982-ApJ,Horne-Baliunas-1986-ApJ,Press-Rybicki-1989-ApJ,Zhou-Sornette-2002-IJMPC},
which is equivalent to the conventional spectrum analysis for evenly
sampled time series. The normalized Lomb power is defined as follows
\begin{equation}\label{Eq:PN}
    P_N(f)=\frac{1}{2\sigma^2} \left\{\frac{[\sum_{j}{(y_j-\overline{y})\cos\omega(t_j-\tau)}]^2}{\sum_{j}{\cos^2\omega(t_j-\tau)}}+\frac{[\sum_{j}{(y_j-\overline{y})\sin\omega(t_j-\tau)}]^2}{\sum_{j}{\sin^2\omega(t_j-\tau)}}\right\}~,
\end{equation}
where $y=S_i$ is the 30-second spread time series of size $N$,
$\omega=2\pi{f}$, $\bar{y}$ and $\sigma$ are the mean and standard
deviation of $y$ (or $S_i$), and
\begin{equation}
    \tan(2\omega\tau)=\frac{\sum_{j}\sin(2\omega t_j)}{\sum_{j}\cos(2\omega
    t_j)}~,
\end{equation}
which determines the parameter $\tau$.

\begin{figure}[htb]
\centering
\includegraphics[width=6.5cm]{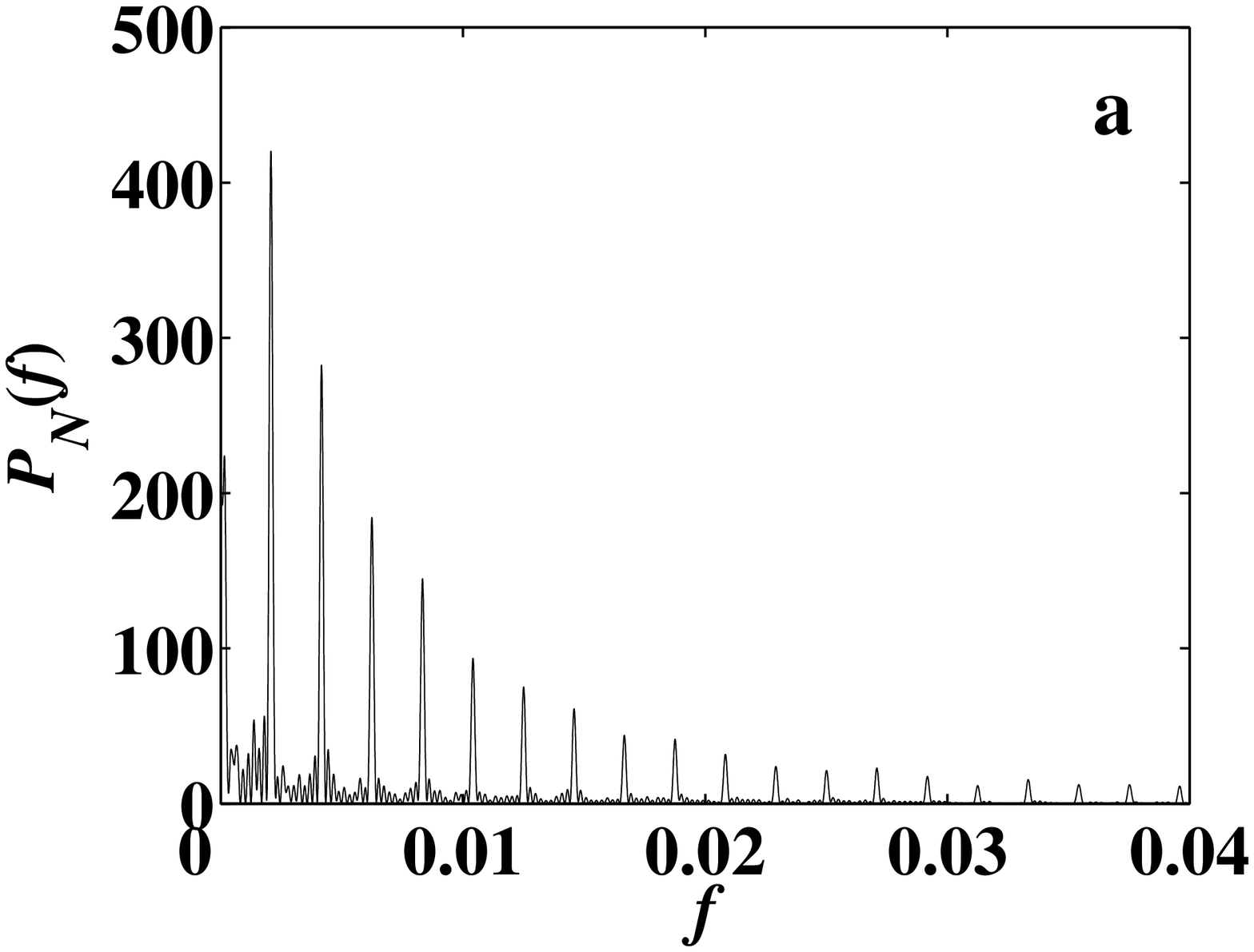}
\includegraphics[width=6.5cm]{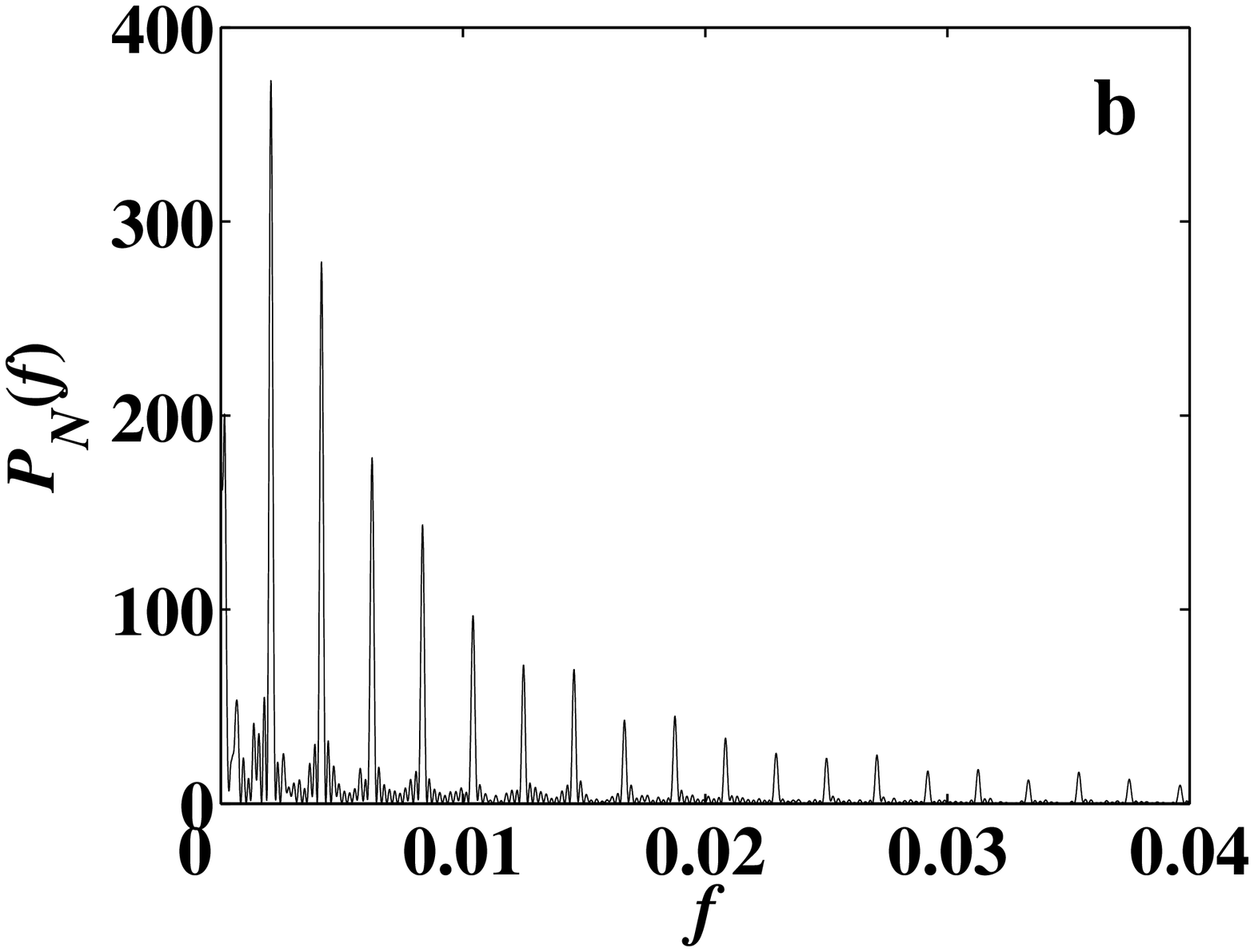}
\caption{Normalized Lomb power $P_N(f)$ for the market averaged
spread time series $S(t')$ of stocks traded on the SHSE (a) and SZSE
(b). } \label{Fig:Lomb}
\end{figure}

Figure \ref{Fig:Lomb} illustrates the Lomb periodograms of the
bid-ask spread time series $S(t')$ for the two Chinese stock
markets. The highest Lomb peak in Fig.~\ref{Fig:Lomb}(a) for SHSE
stocks has height $P_N=420.4$ located at $f=0.002073$, whose
$p$-value is $\ll0.0001$
\cite{Horne-Baliunas-1986-ApJ,Press-Rybicki-1989-ApJ,Zhou-Sornette-2002-IJMPC}.
This gives a period of $1/f=482.4$ data points corresponding
approximately to one trading day. The highest Lomb peak in
Fig.~\ref{Fig:Lomb}(b) for SZSE stocks has height $P_N=372.6$
located at $f=0.002073$, whose $p$-value is $\ll0.0001$. This also
gives a period of roughly one trading day.

\begin{figure}[htb]
\centering
\includegraphics[width=8cm]{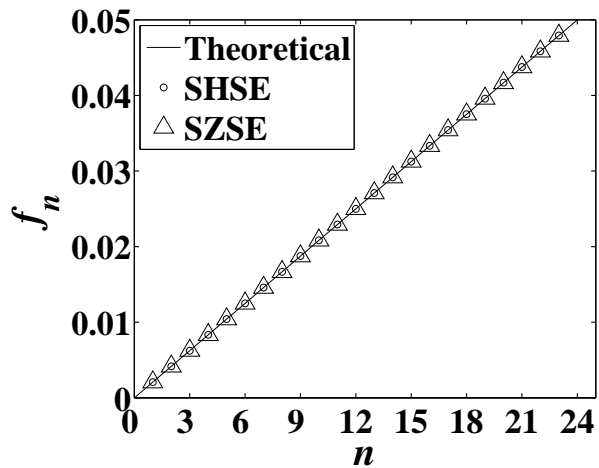}
\caption{Determination of the fundamental frequency $f_0$ with a
linear fit: the open circles are all peaks indicated in
Fig.~\ref{Fig:Lomb}(a) and the open upward triangles are those peaks
in Fig.~\ref{Fig:Lomb}(b). Excellent linear fits are obtained with
the slopes  $f_0 = 0.0020840\pm0.0000003$ ($\circ$) and
$0.0020837\pm0.0000003$ ($\vartriangle$), close to the theoretical
value $f=1/480 = 0.0020833$. The solid line is the theoretical
prediction $f_n=n/480$.} \label{Fig:fn:n}
\end{figure}

A crucial feature observed in Fig.~\ref{Fig:Lomb} is the presence of
harmonic peaks evenly spaced, which provides further evidence for
the presence of periodicity and strengthens its statistical
significance. These harmonics peaks enable us to estimate the
fundamental frequency alternatively in a more accurate manner
\cite{Zhou-Sornette-2002-PD,Zhou-Sornette-Pisarenko-2003-IJMPC,Zhou-Sornette-2004-XXX}.
The approach applies the simple relationship between the fundamental
frequency $f_0$ and its harmonics $f_n$
\begin{equation}
 f_n=n \times f_0~.
 \label{Eq:fn}
\end{equation}
According to Fig.~\ref{Fig:Lomb}, the values of $f_n$ with
$n=1,2,\cdots,23$ are 0.002073, 0.004162, 0.006236, 0.00833,
0.01042, 0.01250, 0.01459, 0.01666, 0.01876, 0.02084, 0.02293,
0.02499, 0.02708, 0.02917, 0.03125, 0.03334, 0.03542, 0.03752,
0.03957, 0.04169, 0.04376, 0.04583, 0.04793 for the SHSE market
averaged spreads, and 0.002073, 0.004162, 0.006251, 0.00833,
0.01042, 0.01250, 0.01459, 0.01667, 0.01875, 0.02084, 0.02293,
0.02501, 0.02709, 0.02919, 0.03127, 0.03335, 0.03542, 0.03751,
0.03960, 0.04165, 0.04376, 0.04584, 0.04791 for the SZSE market
averaged spreads. In order to estimate as accurately as possible the
numerical value of the fundamental frequency $f_0$,
Fig.~\ref{Fig:fn:n} plots the frequencies $f_n$ of all the peaks
found in Fig.~\ref{Fig:Lomb} as a function of the peak sequence
number $n$. Least-squares linear regression of the 23 harmonics
gives $f_0 = 0.0020840\pm0.0000003$ for the SHSE case and
$f_0=0.0020837\pm0.0000003$ for the SZSE case. These values are very
close to the theoretic ansatz $f =1/480 \approx 0.0020833$. We are
thus able to conclude that there is evident daily periodicity in the
market averaged spread $S(t')$. This property holds for the spread
$S_i(t')$ of individual stocks.

\subsection{The intraday patterns}
\label{s2:BAS:intraday}

Based on the daily periodicity in $S_i(t')$ and $S(t')$, we are able
to study the intraday patterns in the spread. Specifically, we
average the spread with same intraday time over different trading
days, which gives
\begin{subequations}
\begin{equation}
  \overline{S}_i(\tau)=\frac{1}{D_i}\sum_{d=1}^{D_i}{S_i(480(d-1)+\tau)}~,~~~\tau=1,\cdots,480
  \label{Eq:Si:bar}
\end{equation}
for individual stock $i$, where $D_i$ is the number of trading days
of stock $i$, and
\begin{equation}
  \overline{S}(\tau)=\frac{1}{D}\sum_{d=1}^D{S(480(d-1)+\tau)}~,~~~\tau=1,\cdots,480
  \label{Eq:S:bar}
\end{equation}
\end{subequations}
for market-average spread, where $D=239$ for the SHSE and $D=240$
for the SZSE.

\begin{figure}[htb]
\centering
\includegraphics[width=6.5cm]{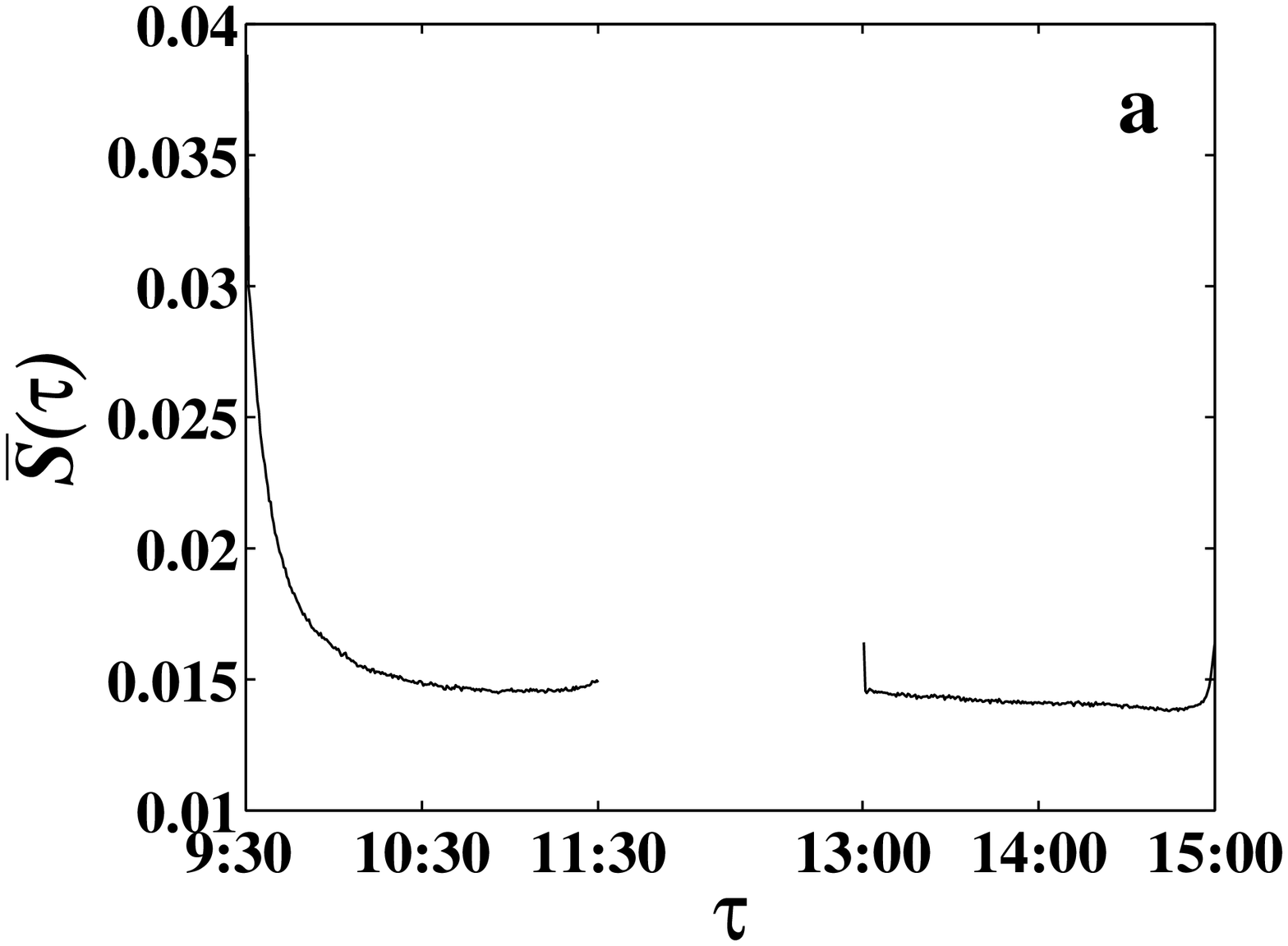}
\includegraphics[width=6.5cm]{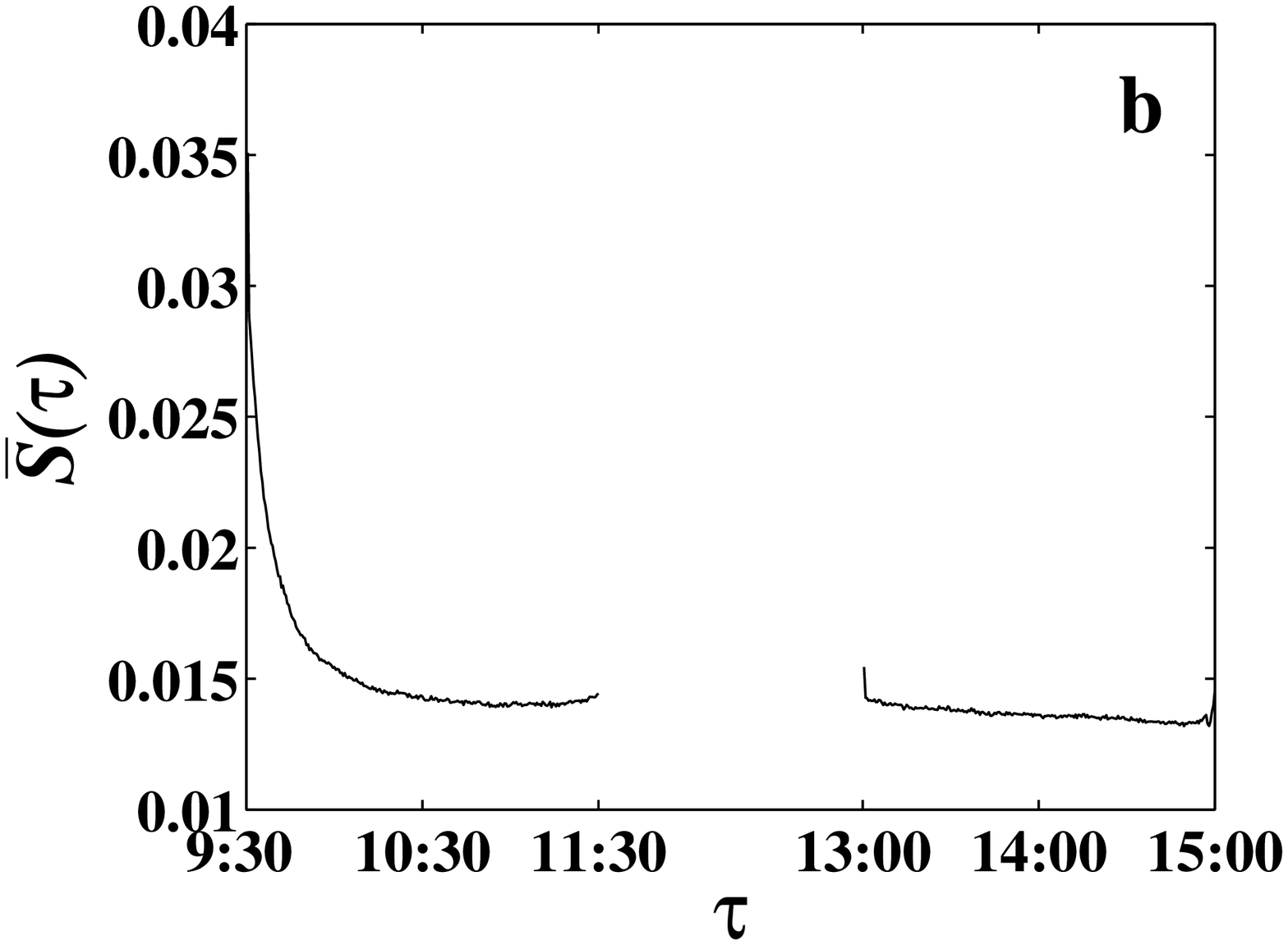}
\caption{Intraday pattern in the bid-ask spreads for all the
A-shares traded on the SHSE (a) and the SZSE (b) in the year 2005.}
\label{Fig:S:bar}
\end{figure}

Figure \ref{Fig:S:bar} shows the intraday pattern in the bid-ask
spreads $\overline{S}(\tau)$ for all the A-shares traded on the SHSE
and the SZSE in the year 2005. It is observed that the bid-ask
spreads exhibit $L$-shaped intraday pattern. In general, the spread
is the largest close to 0.04 CNY (Chinese Yuan) at the open of the
continuous double auction and then declines fast to the average
level of 0.015 CNY an hour later. Before the closing of the markets
at 11:30AM, the spread increases slightly. When the markets reopen
at 13:00PM, the spread is again wider than average, which offsets to
the normal level in a few minutes. The markets close with a slight
increase in the spread at 15:00PM. The situation for individual
stocks is qualitatively similar, especially between 9:30AM and
10:30AM. This well established pattern allows us to investigate in
detail the relaxation behavior of spreads during this time period.

\section{Power-law relaxation of intraday spreads}
\label{s1:PLs}

\subsection{Power-law decay of intraday spreads $\overline{S}(\tau)$}
\label{s2:PL1}

\begin{figure}[htb]
\centering
\includegraphics[width=6.5cm]{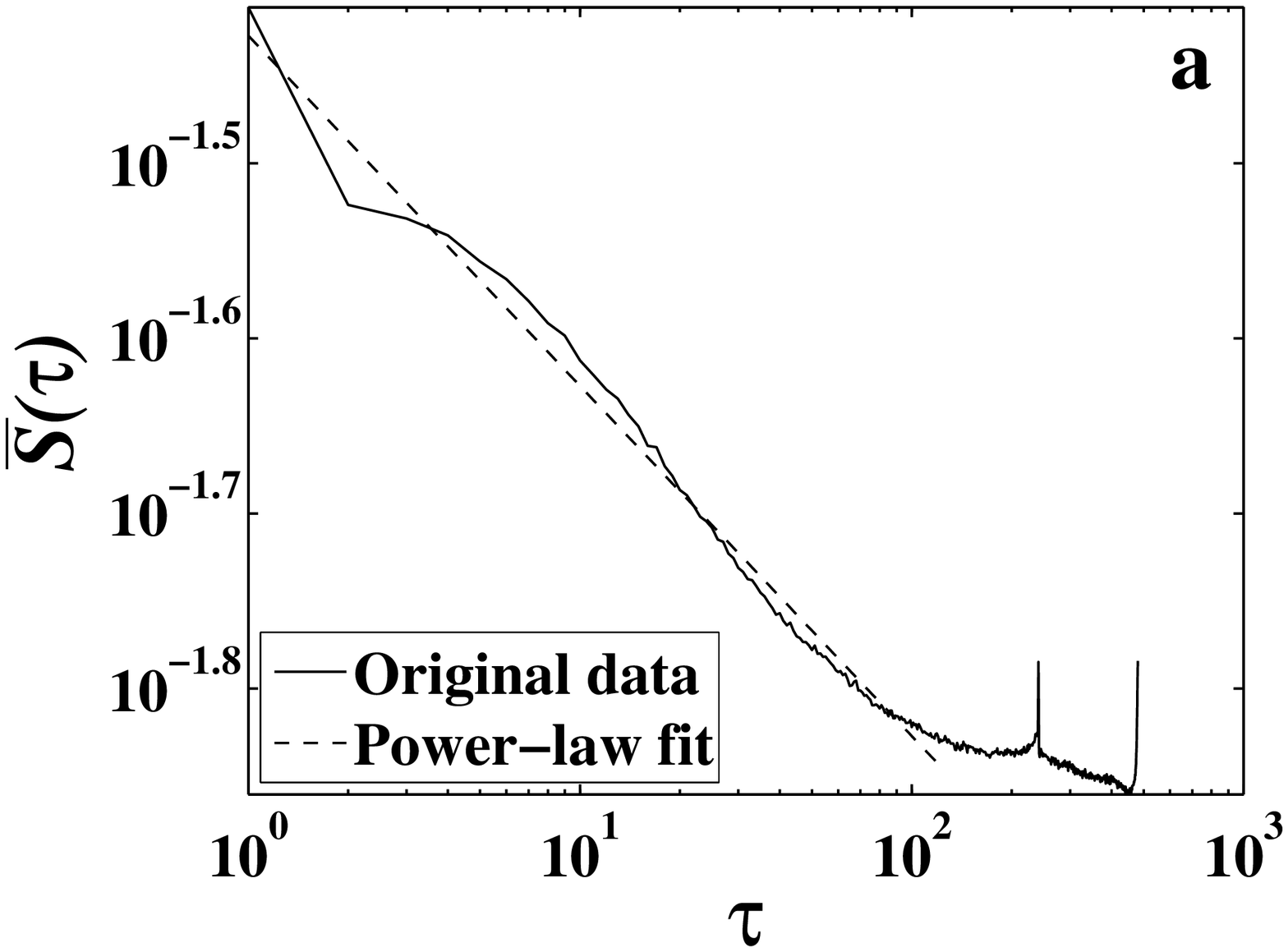}
\includegraphics[width=6.5cm]{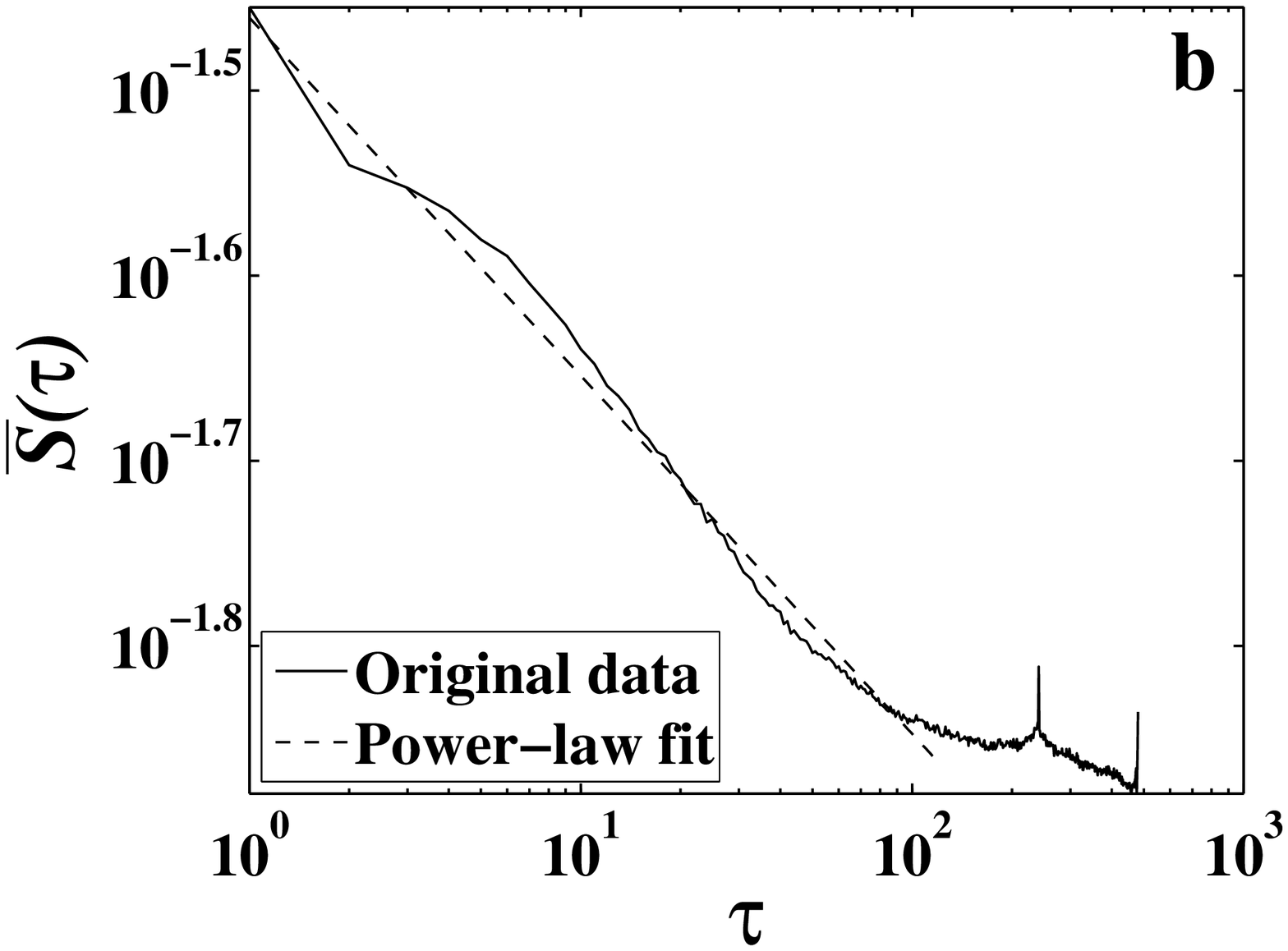}
\caption{Power-law relaxation of $\overline{S}(\tau)$ with respect
to $\tau$ within the first hour of continuous double auction from
9:30AM to 10:30AM for SHSE stocks (a) and SZSE stocks (b). The
dashed lines are the best power-law fits in the scaling range
$\tau\in[1,120]$. The unit of $\tau$ is 30 seconds.}
\label{Fig:PL:Markets}
\end{figure}

Figure \ref{Fig:PL:Markets} shows the averaged intraday spread
$\overline{S}(\tau)$ as a function of $\tau$ in double logarithmic
coordinates for the SHSE and SZSE. We find that the spread
$\overline{S}(\tau)$ decays roughly as a power law in the first hour
from 9:30AM to 10:30AM ($0<\tau\leqslant120$) followed by two sharp
spikes, one at 13:00PM ($\tau=240$) and the other at 15:00PM
($\tau=480$). The relaxation of intraday spread in the first hour
can be expressed as follows
\begin{equation}\label{Eq:PL:Markets}
      \overline{S}(\tau) \sim \tau^{-\beta}~,
\end{equation}
where $\beta=\beta_{\rm{SHSE}}=0.20\pm0.002$ for the SHSE stocks and
$\beta=\beta_{\rm{SZSE}}=0.19\pm0.002$ for the SZSE stocks. In order
to test the relativity between $\tau$ and $S$ in the regression
equations, $t$-test is used here. The $t$-statistic for SHSE is
$|T|=93.025\gg4.0277=t_{0.99995}(118)$, while that for SZSE is
$|T|=83.636\gg4.0277=t_{0.99995}(118)$.

\subsection{Power-law decay of intraday spreads $\overline{S}_i(\tau)$}
\label{s2:PL2}

\begin{figure}[htb]
\centering
\includegraphics[width=6.5cm]{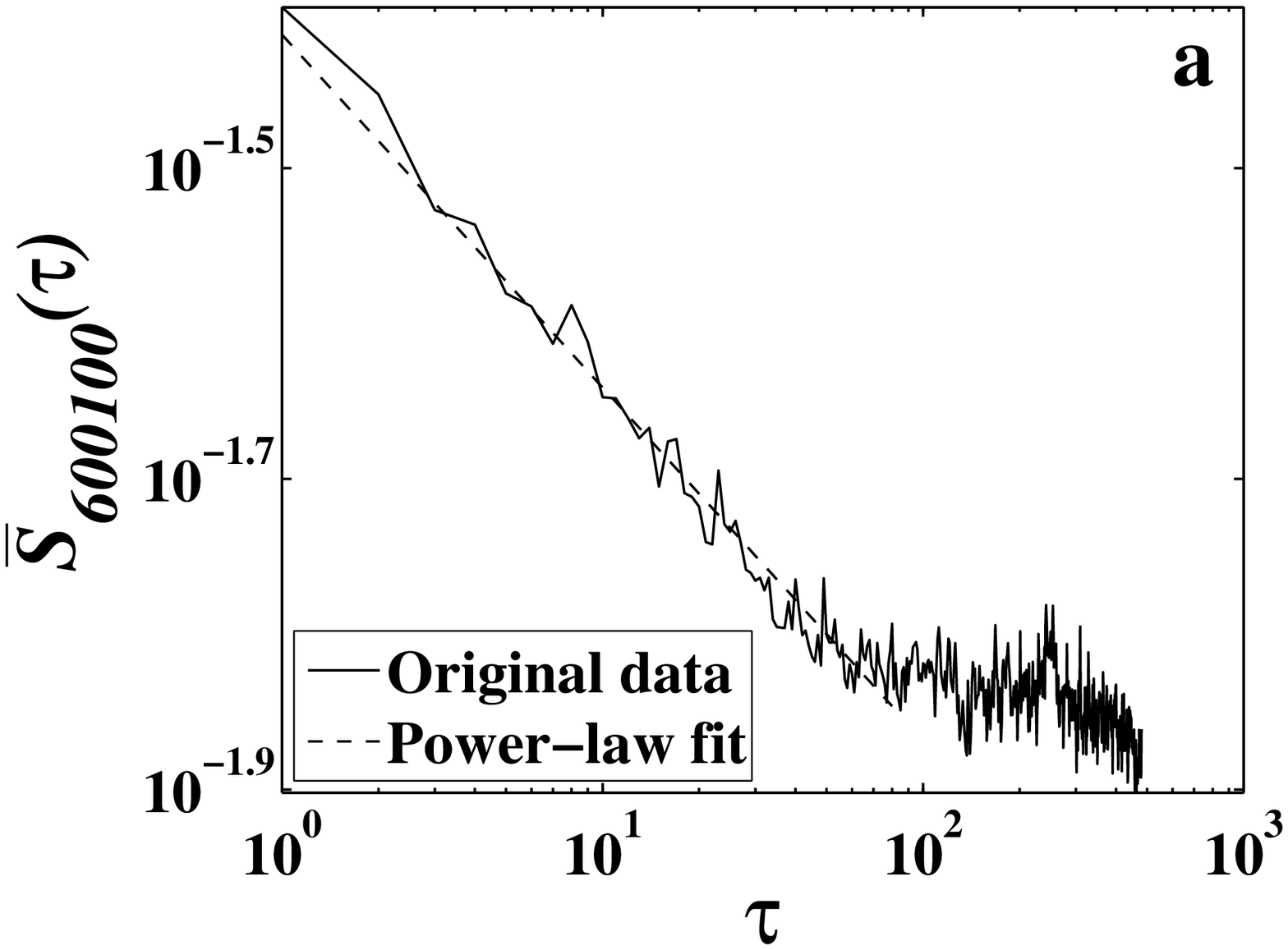}
\includegraphics[width=6.5cm]{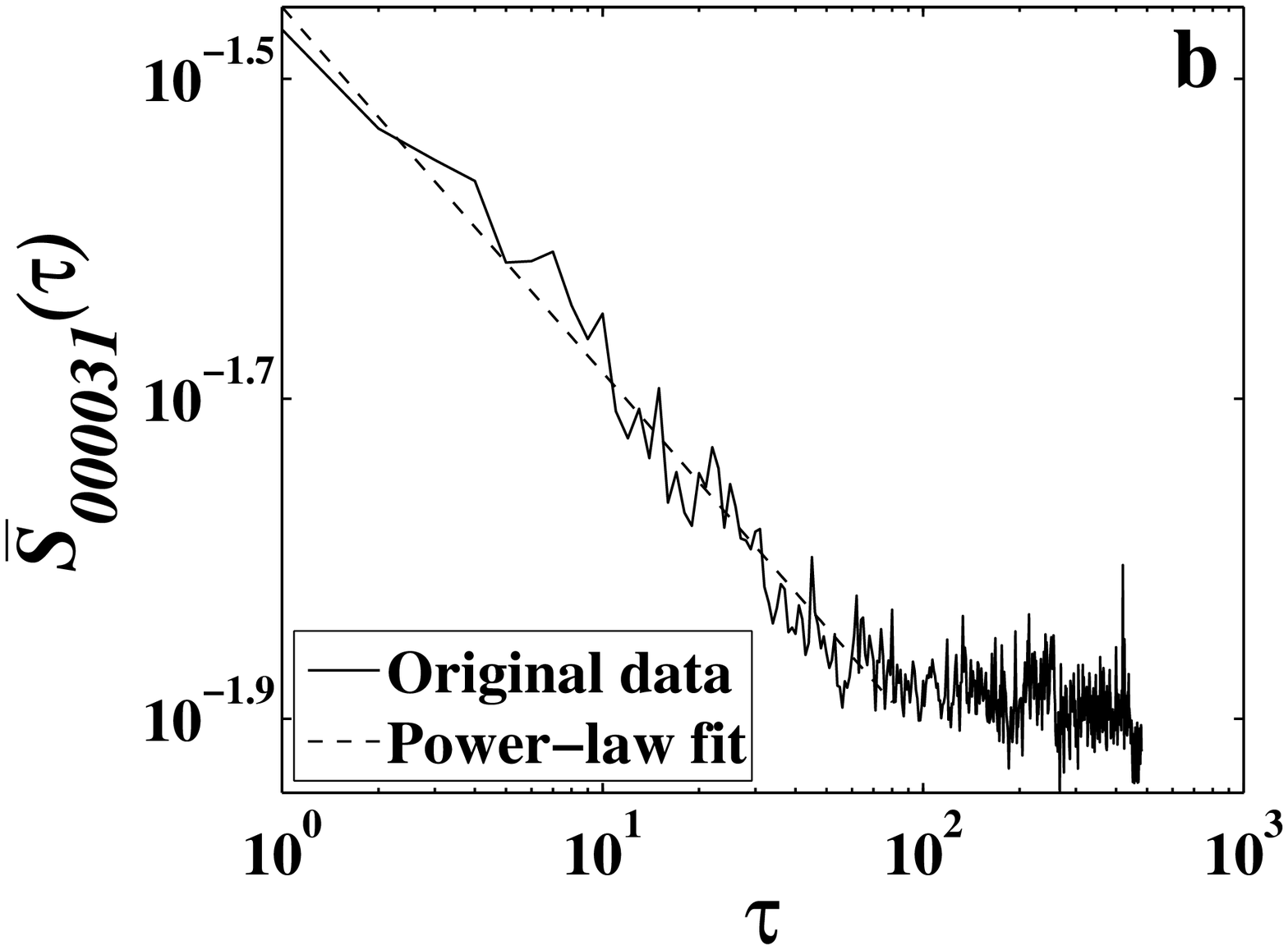}
\caption{Power-law relaxation of $\overline{S}_i(\tau)$ with respect
to $\tau$ within the first 40 minutes of continuous double auction
from 9:30AM to 10:10AM for SHSE stock 600100 (a) and SZSE stock
000031 (b). The dashed lines are the best power-law fits in the
scaling range $\tau\in[1,80]$.} \label{Fig:PL:Stocks}
\end{figure}

When the intraday spreads $\overline{S}_i(\tau)$ for individual
stocks are considered, we find similar power-law relaxation in the
first hour of the continuous double auction. Two typical examples
(stock 600100 traded on the SHSE and 000031 traded on the SZSE) are
illustrated in Fig.~\ref{Fig:PL:Stocks}. The power-law behavior can
be fitted to the following formula
\begin{equation}\label{Eq:PL:Stocks}
      \overline{S}_i(\tau) \sim \tau^{-\beta_i}~.
\end{equation}
Least-squares regressions give $\beta_{\rm{600100}}=0.23\pm0.005$
and $\beta_{\rm{000031}}=0.23\pm0.006$. The scaling range used for
fitting is $\tau\in[1,80]$. The relaxation exponents of these two
arbitrarily chosen stocks are not completely close to the
market-averaged exponents. This calls for a detailed investigation
of the distribution of the power-law relaxation exponents $\beta_i$.

\subsection{Distribution of power-law exponents}
\label{s2:PL3}

\begin{figure}[htb]
\centering
\includegraphics[width=6.5cm]{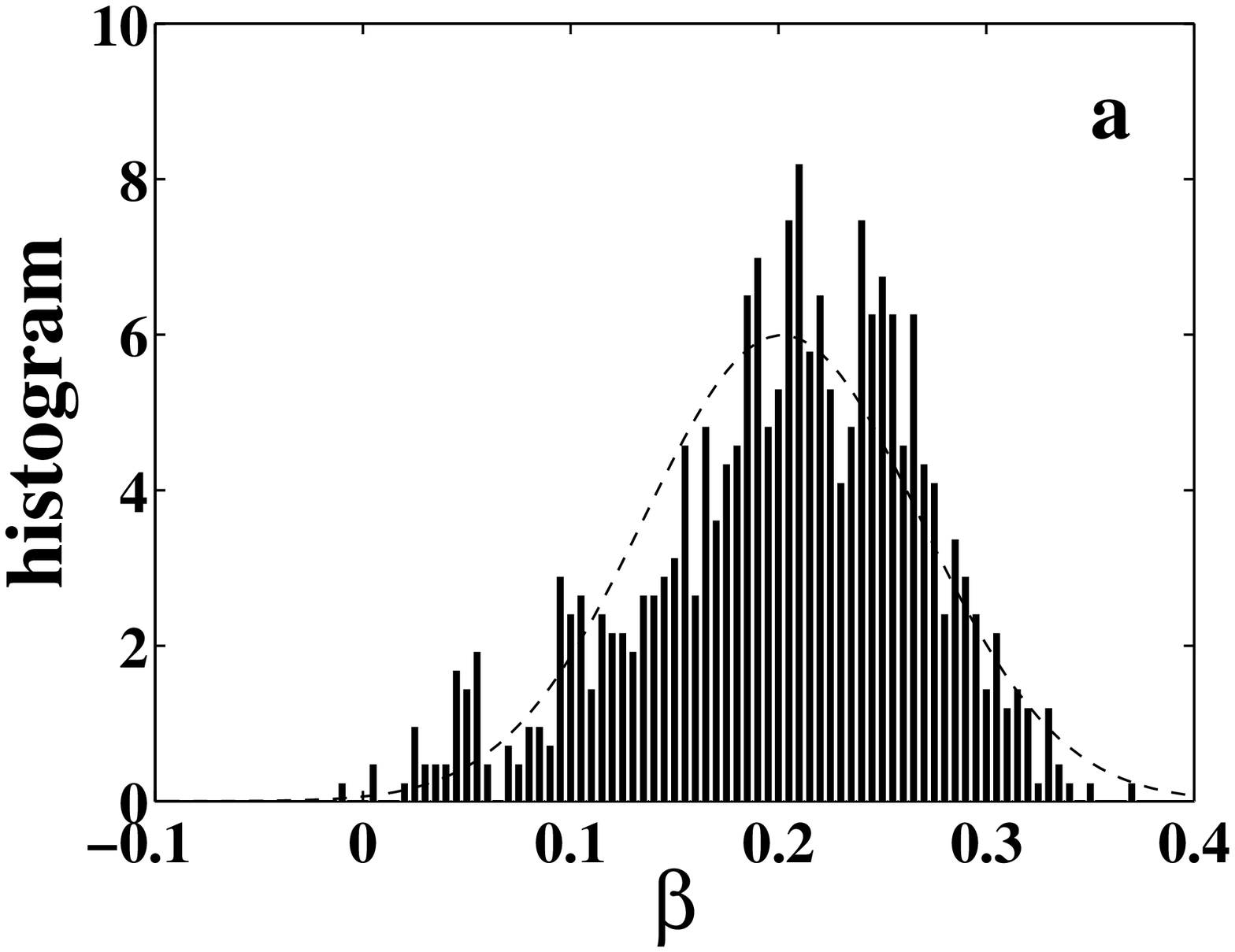}
\includegraphics[width=6.5cm]{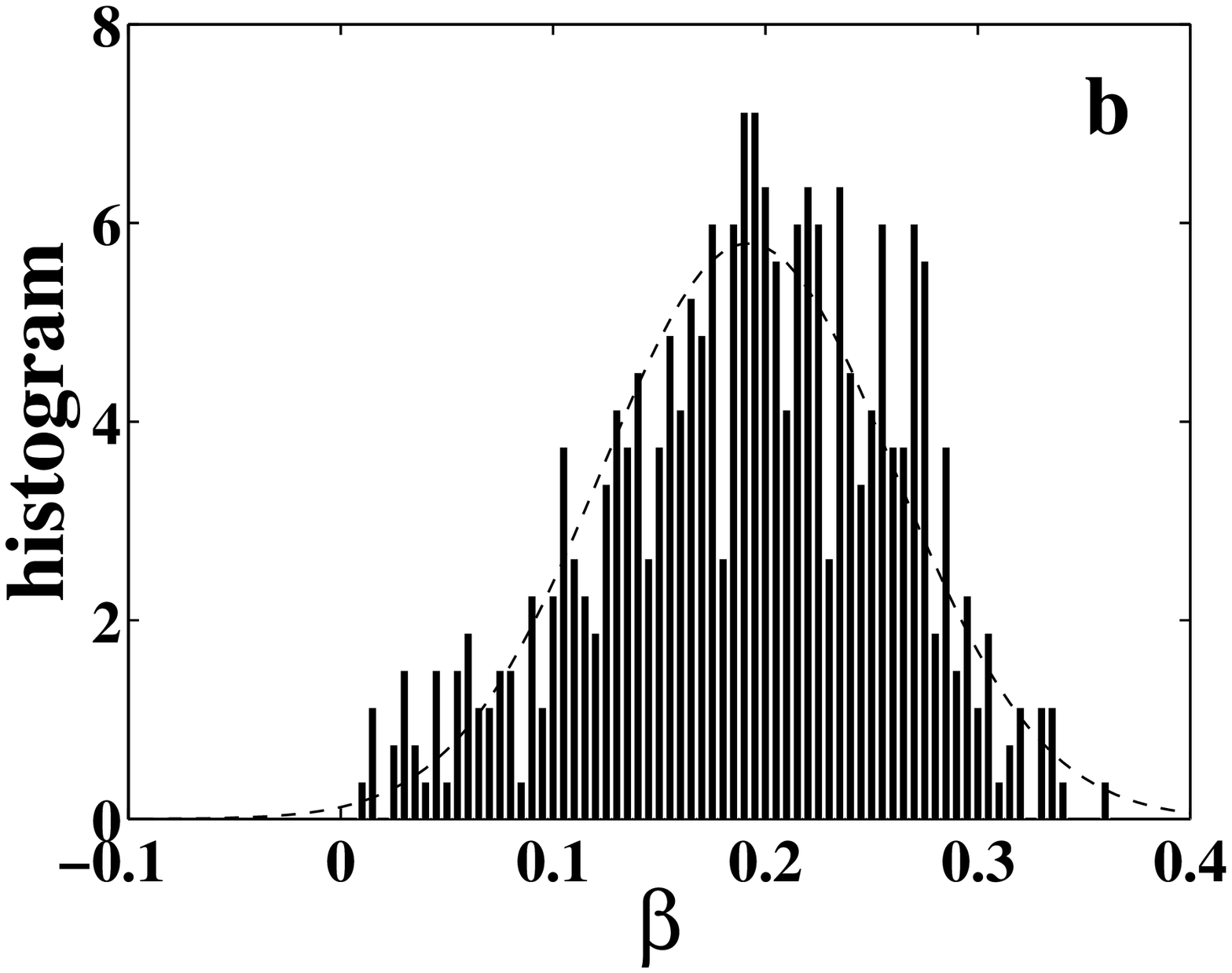}
\caption{Histogram of relaxation exponent $\beta$ for all the stocks
traded on the SHSE (a) and the SZSE (b). The dashed lines in both
plots are the best fits to Gaussian distributions.}
\label{Fig:Hist:beta}
\end{figure}

Here we investigate in detail the relaxation exponent $\beta_i$ of
intraday spread $\overline{S}_i(\tau)$ for individual stocks traded
on the SHSE and the SZSE. Figure \ref{Fig:Hist:beta} draws the
histograms of the power-law relaxation exponent $\beta$ for the SHSE
and the SZSE respectively. We find that
$\beta_{\rm{SHSE}}=0.20\pm0.067$ and
$\beta_{\rm{SZSE}}=0.19\pm0.069$. In addition, the skewness are
$\nu_{\rm{SHSE}}=-0.5330$ and $\nu_{\rm{SZSE}}=-0.3309$, while the
kurtosis are $\kappa_{\rm{SHSE}}=3.0187$ and
$\kappa_{\rm{SZSE}}=2.7016$. The two distributions can be modeled
roughly by normal distributions. Since the mean and variance of the
sample are unbiased estimates for normal distribution, the null
hypothesis $H_0$ can be expressed as follows
\begin{equation}
 \left\{
 \begin{array}{ccc}
 \beta_{\rm{SHSE}} \sim \mathcal{N}(0.20,0.0067)\\
 \beta_{\rm{SZSE}} \sim \mathcal{N}(0.19,0.0069)
 \end{array}\right.~,
 \label{Eq:Null}
\end{equation}
where $\mathcal{N}(\mu, \sigma^2)$ is the normal distribution with
mean $\mu$ and variance $\sigma^2$. We test the hypothesis by means
of $\chi^2$ test. We find that
$\chi^2_{\rm{SHSE}}=155<157=\chi^2_{0.9999}(97)$,
$\chi^2_{\rm{SZSE}}=95\leqslant96=\chi^2_{0.5}(97)$. Speaking
differently, the normality of power-law relaxation exponents $\beta$
for both SHSE and SZSE stocks can not be rejected.

\section{Summary}
\label{s1:conclusion}

We have studied the intraday patterns in the bid-ask spreads of 1364
Chinese A-share stocks using high-frequency data. The daily
periodicity in the spread time series is confirmed by Lomb analysis.
The intraday bid-ask spreads exhibit $L$-shaped pattern. The
intraday spread of individual stocks relaxes as a power-law in the
first 40 minutes of the continuous double auction from 9:30AM to
10:10AM. We have found that the averaged power-law decay exponents
are $\beta_{\rm{SHSE}}=0.20\pm0.067$ and
$\beta_{\rm{SZSE}}=0.19\pm0.069$. The chi-squared test shows that
the relaxation exponent $\beta$ of individual stocks is normally
distributed.

It is interesting to note that the value of $\beta$ enables us to
classify the widening of the spread in the cooling period (from
9:25AM to 9:30AM) as an endogenous process. According to the network
theory of peaks in complex systems
\cite{Sornette-Deschatres-Gilbert-Ageon-2004-PRL}, an endogenous
peak decays as a power law $\sim \tau^{1-2\theta}$, while an
exogenous peak relaxes as $\sim \tau^{1-\theta}$. We find that
$\theta \approx 0.4$, which is in line with the endogenous exponents
observed in other complex systems such as the dynamics of book sales
and online download of articles
\cite{Sornette-Deschatres-Gilbert-Ageon-2004-PRL,Deschatres-Sornette-2005-PRE,Johansen-Sornette-2000-PA,Johansen-2001-PA}.

\bigskip
{\textbf{Acknowledgments:}}

We are grateful to Gao-Feng Gu for fruitful discussion and Liang Guo
for preprocessing the raw data. This work was partly supported by
the National Natural Science Foundation of China (Grant No.
70501011), the Fok Ying Tong Education Foundation (Grant No.
101086), and the Shanghai Rising-Star Program (Grant No. 06QA14015).

%\pagebreak
\bibliography{E:/Papers/Auxiliary/Bibliography} %Zhou
%\bibliography{Bibliography}
%\bibliography{F:/Bibliography}

\end{document}